%% file: main.tex
\pdfoutput=1
\documentclass[10pt,twocolumn,letterpaper]{article}

\usepackage[pagenumbers]{iccv} % To force page numbers, e.g. for an arXiv version

\usepackage[utf8]{inputenc}
\usepackage[T1]{fontenc}
\definecolor{iccvblue}{rgb}{0.21,0.49,0.74}
\usepackage[pagebackref,breaklinks,colorlinks,allcolors=iccvblue]{hyperref}
\usepackage{graphicx}
\usepackage[section]{placeins}
\usepackage{tabularx}
\usepackage{multirow}
\usepackage{CJKutf8}
\usepackage{amsmath} % for the equation* environment
\usepackage{caption}
\usepackage{float}
\usepackage{ulem}
\def\paperID{9333} % *** Enter the Paper ID here
\def\confName{ICCV}
\def\confYear{2025}
\usepackage{adjustbox}
\title{FuXi-RTM: A Physics-Guided Prediction Framework\\ with Radiative Transfer Modeling}
\author{$\text{Qiusheng Huang}^{1,2,3}$\textsuperscript{\dag}, $\text{Xiaohui Zhong}^{1,3}$\textsuperscript{\dag}, $\text{Xu Fan}^3$, $\text{Lei Chen}^{1,3}$, $\text{Hao Li}^{1,2,3}$\footnotemark[1]\\
$^1$Artificial Intelligence Innovation and Incubation Institute, Fudan University\\
$^2$Shanghai Innovation Institute\\
$^3$Shanghai Academy of Artificial Intelligence for Science\\
}

\begin{document}
\maketitle
\renewcommand{\thefootnote}{\fnsymbol{footnote}}
\footnotetext[1]{Corresponding author.}
\footnotetext[2]{These authors contributed equally to this work.}
\renewcommand{\thefootnote}{\arabic{footnote}}
\begin{abstract}
Similar to conventional video generation, current deep learning-based weather prediction frameworks often lack explicit physical constraints, leading to unphysical outputs that limit their reliability for operational forecasting. Among various physical processes requiring proper representation, radiation plays a fundamental role as it drives Earth's weather and climate systems. However, accurate simulation of radiative transfer processes remains challenging for traditional numerical weather prediction (NWP) models due to their inherent complexity and high computational costs. Here, we propose FuXi-RTM, a hybrid physics-guided deep learning framework designed to enhance weather forecast accuracy while enforcing physical consistency. FuXi-RTM integrates a primary forecasting model (FuXi) with a fixed deep learning-based radiative transfer model (DLRTM) surrogate that efficiently replaces conventional radiation parameterization schemes. This represents the first deep learning-based weather forecasting framework to explicitly incorporate physical process modeling. Evaluated over a comprehensive 5-year dataset, FuXi-RTM outperforms its unconstrained counterpart in 88.51\% of 3320 variable and lead time combinations, with improvements in radiative flux predictions. By incorporating additional physical processes, FuXi-RTM paves the way for next-generation weather forecasting systems that are both accurate and physically consistent.\\
% \noindent\footnotetext[1]{\textsuperscript{\dag}These authors contributed equally to this work.}
% \noindent\footnotetext[1]{$*$Corresponding author.}
%Accurate simulating radiation is essential for weather and climate forecasting, yet traditional numerical weather prediction (NWP) models face significant challenges due to the complexity of radiative transfer processes and high computational costs. However, similar to conventional RGB video generation, deep learning-based weather prediction frameworks often lack explicit physical constraints, leading to unphysical outputs. Here we propose FuXi-RTM, a hybrid physics-guided deep learning framework designed to enhance weather forecast accuracy and physical consistency while extending predictions for radiation. FuXi-RTM integrates a trainable primary forecasting model (FuXi) with a deep learning-based radiative transfer model (DLRTM) surrogate, which replaces conventional radiation parameterization schemes. This represents the first deep learning-based weather forecasting framework to explicitly model physical processes. Evaluated over a 5-year dataset, FuXi-RTM outperforms its unconstrained counterpart in 88.51\% of 3320 variable and lead time combinations, with more significant improvements in radiative flux predictions. By incorporating additional physical processes, FuXi-RTM paves the way for next-generation weather forecasting systems that are both accurate and physically consistent.
\end{abstract}

\section{Introduction}
\label{sec:intro} % Accurate Precisely
Accurate modeling of multi-channel spatiotemporal sequences with complex physical constraints presents significant challenges beyond conventional video prediction tasks. While both tasks essentially predict the next state of physical systems, RGB video prediction primarily optimizes for perceptual quality and visual coherence, whereas physical system forecasting - exemplified by weather prediction - demands consistency across multidimensional, physically interacting variables. Unlike RGB videos with three fixed channels, weather forecasting involves dozens of interrelated variables across multiple atmospheric layers, governed by physical laws rather than mere visual coherence—requiring models that maintain consistency across all prediction channels simultaneously. To address this fundamental challenge, the most intuitive approach is to incorporate strong physical process constraints that explicitly model the underlying dynamics governing these complex systems, ensuring predictions remain consistent with established physical laws. Among various physical processes, radiative transfer presents an ideal starting point for such integration, given its well-defined physics and fundamental importance.
Indeed, radiation is the primary source of energy that drives the Earth's weather and climate systems, modulating temperature gradients, atmospheric pressure patterns, wind circulations, and moisture distribution \cite{Kevin2009,Stephens2012}.
Therefore, accurately simulating radiative processes is essential for weather and climate forecasting, which has significant implications for socioeconomic planning and daily human activities.
Numerical weather prediction (NWP) models, which have been foundational tools in meteorology since their emergence in the 1950s \cite{Charney1950,pu2019numerical}, simulate radiative transfer processes by resolving interactions between shortwave (SW) and longwave (LW) radiation and atmospheric constituents (e.g., clouds, water vapor, and aerosols), as well as the Earth's surface \cite{Ritter1992}.
The accuracy of these models depends on spatial resolution, parameterization schemes \cite{stensrud2009parameterization}, and the quality of initial conditions.
While advancements in NWP models have steadily enhanced forecast accuracy over recent decades \cite{bauer2015quiet}, challenges persist due to the inherent complexity of cloud microphysics and radiative transfer processes \cite{rad2017}.
These challenges are particularly pronounced under cloudy conditions, where uncertainties in cloud properties and vertical structures propagate errors in simulated radiative fluxes \cite{Tuononen2019}.
Further enhancement of conventional physics-based models faces increasing computational barriers.
In contrast, deep learning is revolutionizing weather prediction \cite{bouallegue2024rise}, with data-driven models demonstrating superior computational efficiency and forecast skill for conventional meteorological variables (e.g., temperature, wind, and pressure) compared to the high-resolution deterministic forecasts (HRES) by the European Center for Medium-Range Weather Forecasts (ECMWF) \cite{ECMWF2021}, the world's leading operational prediction center \cite{pathak2022fourcastnet,bi2022panguweather,lam2022graphcast,chen2023fuxi,chen2023fengwu,chen2023fuxis2s,lang2024,nguyen2025scaling,yuan2025tianxing}.
However, current deep learning-based forecasting models remain fundamentally physics-agnostic and may produce unphysical outputs, such as negative humidity \cite{schreck2024}.
This situation
%lack of explicit physical constraints 
raises concerns about the physical plausibility and long-term stability of these predictions \cite{wattmeyer2023a}, particularly for radiative processes, which are underexplored yet.

%XH，本段主要是描述一些融入物理约束的工作以及相关的不足，尤其是没有融入物理参数化方案的约束
To date, the integration of rigid physical processes constraints into deep learning-based weather forecasting models has not been realized.
%XH，这里需要概括技术创新点，这里的只提到了一些idea，但是没有一些可以细说的技术点
In this study, we propose FuXi-RTM, a hybrid physics-guided deep learning architecture that is designed to integrate data-driven weather forecasts with physics-aware constraints.
FuXi-RTM consists of two main components: 1) a primary proven forecasting model based on FuXi \cite{chen2023fuxi}, and 2) a deep learning-based radiative transfer model (DLRTM).
This hybrid design combining the flexibility of data-driven models with weather domain-specific physics, enhancing the physical plausibility and accuracy of the forecasting model’s outputs while maintaining computational efficiency.
Experimental results show that FuXi-RTM outperforms its unconstrained counterpart on 88.51\% of 3320 variable and lead time combinations (and on 100\% for radiative fluxes).
By extending this architecture to incorporate other critical physical processes, such as convection, planetary boundary layer (PBL), land surface interactions, and cloud microphysics, deep learning-based weather systems can achieve unprecedented accuracy and physical consistency. Our key contributions as follows:
\begin{itemize}[leftmargin=2em]
    \item We propose FuXi-RTM, the first physics-guided spatiotemporal sequence prediction that integrates explicit physical process modeling with excellent weather forecasting capabilities.
    \item We propose a straightforward yet effective framework that extends weather prediction models with radiative transfer capabilities without additional training, while achieving orders of magnitude improvement in computational efficiency over traditional schemes. 
    \item We present comprehensive experiments demonstrating that our approach not only improves forecast accuracy across meteorological variables and radiation fluxes but also significantly enhances physical consistency, as validated through physical conservation evaluations.
\end{itemize}
\section{Related work}
\subsection{Video Generation and Prediction Models}
Video Generation research has evolved rapidly, focusing on generating coherent spatiotemporal sequences for tasks ranging from human motion forecasting to natural scene evolution. Recent advances in diffusion-based models \cite{blattmann2023align, ho2022video, khachatryan2023text2video, luo2023videofusion, singer2022make, wang2023videofactory, xing2024make, videoworldsimulators2024, opensora} have demonstrated impressive results in capturing visual coherence and perceptual quality. However, these approaches face fundamental limitations when applied to atmospheric science due to weather data's unique characteristics: high-resolution multi-variable structure creates prohibitive memory requirements, while the physical interdependence among diverse atmospheric variables necessitates expensive retraining of representation models(like VAEs \cite{kingma2013auto, van2017neural}) whenever variable combinations change. 
\subsection{Weather Forecast}
To address these challenges, weather forecasting develops specialized Spatiotemporal architectures, with recent advances in graph-based frameworks \cite{lam2022graphcast, pathak2022fourcastnet} and transformer-based approaches \cite{bi2022panguweather,chen2023fuxi,chen2023fengwu,chen2023fuxis2s,lang2024,nguyen2025scaling,yuan2025tianxing} demonstrating significant improvements. FuXi \cite{chen2023fuxi} , a state-of-the-art model in this domain, prioritizes forecast skill against ground truth observations rather than perceptual quality. Despite their effectiveness, these weather-specific models remain fundamentally physics-agnostic, relying on data correlations without incorporating the physical laws governing atmospheric dynamics—raising concerns about their reliability under extreme conditions or extended prediction horizons.
\subsection{Physics-Guided Prediction}
In the broader field of video generation, several approaches have implicitly captured physical dynamics. Video generation models like WorldDreamer \cite{wang2024worlddreamer}, 
Sora\cite{videoworldsimulators2024} and OpenSora \cite{opensora} learn to generate physically plausible content through large-scale training on multimodal data, while specialized models such as DrivingFusion \cite{li2024drivingdiffusion}and PSLG \cite{li2024physics} incorporate domain knowledge to enhance physical consistency in specific scenarios. However, studies \cite{kang2024far} reveal that current models struggle to abstract universal physical principles from data alone.\\
For weather forecasting, previous studies \cite{liu2024,verma2024,xu2025} have primarily focused on incorporating primitive equations that describe atmospheric motion using ODE solvers \cite{runge1895numerische,biswas2013} . These approaches, however, lack explicit constraints on physical processes like radiative transfer. In operational NWP models, physical processes are represented through parameterization schemes\cite{stensrud2009parameterization}, which approximate complex atmospheric interactions based on simplified physical assumptions.  %%%%
\begin{figure*}[ht]
\centering
\includegraphics[width=0.95\textwidth]{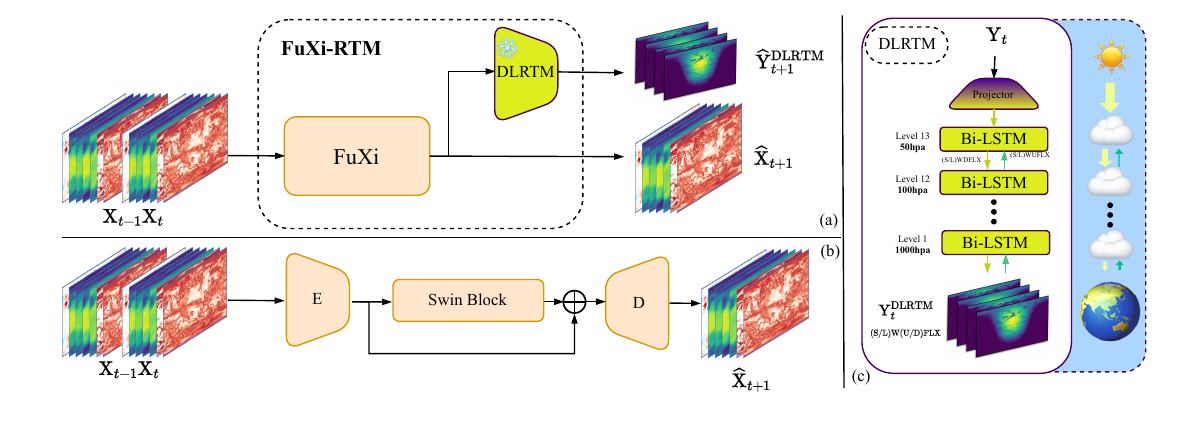}
\caption{\textbf{The overall structure of our method.} (a) Schematic of FuXi-RTM. (b) Architecture of FuXi-base. (c) Schematic of the deep learning-based radiative transfer model (DLRTM) utilizing a bidirectional long short-term memory (Bi-LSTM) architecture.}
\label{FuXi-RTM_model}
\end{figure*}
Recent work has explored using deep learning to emulate these parameterization schemes \cite{Rasp2018,Han2020,yuval2020,yuval2021,YaoZhong2023,Zhong2023,zhong2024machine}, yet integrating such surrogate models with deep learning-based weather forecasting remains an open challenge.
\subsection{Radiative Transfer Modeling}
Atmospheric radiative transfer calculations traditionally rely on line-by-line radiative transfer models (LBLRTM) \cite{clough2005atmospheric,clough1992line}, though their computational costs make them impractical for operational forecasting. Instead, radiation parameterization schemes \cite{Ritter1992} are employed in NWP models, using simplified physics to represent radiation-atmosphere interactions \cite{pu2019numerical}. In the deep learning context, Yao et al. \cite{YaoZhong2023} found that bidirectional LSTM architectures achieve superior accuracy for radiation modeling, providing a foundation for our DLRTM implementation.
\section{Method}
\label{sec:method}
\subsection{Data}
ERA5 \cite{hersbach2020era5}, the fifth iteration of the ECMWF reanalysis dataset, represents the most comprehensive and accurate global reanalysis archive available and is widely used in developing deep learning-based weather forecasting framework.
For this study, we employ the 6-hourly ERA5 dataset with a $0.25^\circ$ resolution (equivalent to $721\times1440$ latitude-longitude grid points). 
Unlike conventional video data with pixels in height (H) and width (W) dimensions and sequential frames, weather data uses grid points representing specific geographic locations defined by latitude and longitude coordinates. Each grid point stores meteorological variables analogous to how pixels store color intensities. Similarly, forecast lead times at 6-hour intervals replace frame indices for temporal progression.
The input to FuXi-RTM consists of four-dimensional cubes $\mathbf{X}_t, \mathbf{X}_{t-1} \in \mathbb{R}^{1 \times \textrm{C} \times \textrm{H} \times \textrm{W}}$, where $\textrm{C}=83$ represents the number of meteorological variables, and $\textrm{H}=721$, $\textrm{W}=1440$ correspond to the spatial dimensions of the global grid. Each such "frame" captures the complete global atmospheric state at time step $t$, analogous to frames in video generation tasks.\\
Specifically, The FuXi-RTM model forecasts 83 variables, including 5 upper-air atmospheric variables across 13 pressure levels (50, 100, 150, 200, 250, 300, 400, 500, 600, 700, 850, 925, and 1000 hPa), and 18 surface variables.
The upper-air atmospheric variables are geopotential (${\textrm{Z}}$), temperature (${\textrm{T}}$), fraction of cloud cover (${\textrm{CC}}$), specific cloud liquid water content (${\textrm{CLWC}}$), and specific humidity (${\textrm{Q}}$).
The surface variables are 2-meter temperature (${\textrm{T2M}}$), 2-meter dewpoint temperature (${\textrm{D2M}}$), 10-meter u wind component (${\textrm{U10M}}$), 10-meter v wind component (${\textrm{V10M}}$), 100-meter u wind component (${\textrm{U100M}}$), 100-meter v wind component (${\textrm{V100M}}$), mean sea-level pressure (${\textrm{MSL}}$), surface pressure (${\textrm{SP}}$), low cloud cover (${\textrm{LCC}}$), medium cloud cover (${\textrm{MCC}}$), high cloud cover (${\textrm{HCC}}$), total cloud cover (${\textrm{TCC}}$), surface albedo (${\textrm{FAL}}$), surface net solar radiation (${\textrm{SSR}}$), surface solar radiation downwards (${\textrm{SSRD}}$), total sky direct solar radiation at surface (${\textrm{FDIR}}$), top net thermal radiation (${\textrm{TTR}}$), and total precipitation (${\textrm{TP}}$).
A complete list of these variables and their abbreviations is provided in Tab. 1 of the supplementary material.
Notably, the u and v components of wind across pressure levels are excluded from the model, as they are not necessary for examining radiative constraints.
% 辐射部分建议加入related work，DLRTM可以在method中说。
% Accordingly, we use ERA5 as input to RRTMG model to generate radiative fluxes across the same 13 pressure levels utilized by FuXi.
\subsection{FuXi model}
Before introducing our proposed FuXi-RTM, we briefly describe the baseline FuXi model architecture. The FuXi model takes as input two consecutive global atmospheric state data cubes $\textrm{X}_t, \textrm{X}_{t-1}$ and aims to predict the atmospheric state at the next time step $\widehat{\textrm{X}}_{t+1}$. 
As an autoregressive model, FuXi extends forecast lead times by recursively feeding its outputs back as inputs. This recursive process continues until reaching the desired forecast horizon (typically 10 days in our experiments).
As shown in Fig.\ref{FuXi-RTM_model}, the architecture follows an encoder-processor-decoder paradigm. First, 3D convolutional layers encode the input data into feature representations. These features are then processed through a backbone network consisting of Swin Transformer V2 blocks\cite{liu2022swin}, which effectively capture long-range dependencies and multi-scale features crucial for global weather systems. Finally, 3D transposed convolutions decode the processed features to generate the predicted atmospheric state. Skip connections between the encoder and decoder preserve detailed information throughout the network. In our implementation, the baseline model FuXi-base employs 30 Swin Transformer V2 blocks, reducing the parameter count from approximately 1.5 billion in the original FuXi to 1.1 billion while maintaining strong predictive capabilities. 
%While FuXi achieves impressive forecast skill, it lacks explicit physical constraints, potentially leading to physically implausible predictions—a limitation we address with our proposed FuXi-RTM model.
\subsection{Deep learning based RRTMG model}
\textbf{Model designs.} Inspired by Yao et al \cite{YaoZhong2023}, we develop a DLRTM based on the Bi-LSTM model.
Our DLRTM consists of three repeated Bi-LSTM layers, each containing a forward and a backward LSTM layer with feature dimensions of 96 and 128, respectively.
As illustrated in Fig.\ref{FuXi-RTM_model}, the DLRTM processes data $\mathbf{Y}_t \in \mathbb{R}^{1 \times 71 \times \textrm{H} \times \textrm{W}}$ selected from $\mathbf{X}_t$, across 13 pressure levels in an atmospheric column, ranging from 50 hPa at the top of the atmosphere (TOA) to 1000 hPa near the surface.
Note that, DLRTM operates independently on each grid point $(i,j)$, processing one vertical atmospheric column at a time. This column-wise approach allows for efficient parallel computation across the global grid.
At each level, the model processes 11 input variables (listed in Tab.2 in the supplementary material), including 5 upper-air variables that vary with pressure levels and 6 single-level variables (e.g. solar zenith angles and land-sea mask) that remain constant across levels.
The DLRTM generates $\mathbf{Y}_t^\mathbf{DLRTM} \in \mathbb{R}^{1 \times \left(4 \times 13\right) \times \textrm{H} \times \textrm{W}}$, including four output variables at each layer: shortwave upward fluxes (SWUFLX), shortwave downward fluxes (SWDFLX), longwave upward fluxes (LWUFLX), and longwave downward fluxes (LWDFLX).

A critical challenge in atmospheric modeling arises from Earth's varying topography. For example, consider a mountainous region like Tibet, where surface pressure might be approximately 600 hPa. In such locations, the standard pressure levels of 700, 850, 925, and 1000 hPa would fall below the actual surface—these non-physical locations are termed "ghost levels" ($P_{\text{level}} > P_{\text{surface}}$). To address this, DLRTM implements dynamic masking:
\begin{equation}
\vspace{-5pt}
\text{M}(i,j,k) = 
\begin{cases}
1 & \text{if}\ P_{\text{level}}(k) \leq P_{\text{surface}}(i,j) \\
0 & \text{if}\ P_{\text{level}}(k) > P_{\text{surface}}(i,j)
\end{cases}
\end{equation}

where $(i,j)$ represents a grid point location and $k$ indexes the pressure level. During forward propagation, DLRTM uses this mask to exclude non-physical levels from calculations, ensuring that radiative fluxes are only computed for atmospheric layers that physically exist.\\
\textbf{Loss designs.} 
To train the DLRTM model effectively, we employ a mean squared error loss function:
\begin{equation}
\label{loss_3}
\resizebox{0.47\textwidth}{!}{$
\textrm{L}_{reg}=
\frac{1}{\textrm{H}\times\textrm{W}\times\textrm{R}}
\sum_{i=1}^{\textrm{H}}
\sum_{j=1}^{\textrm{W}}
\text{M}(i,j) \times
\left(
\sum_{r=1}^{\textrm{R}}
\left((\mathbf{\widehat{Y}^{DLRTM}}_{r,i,j}-\mathbf{Y}^{\mathbf{DLRTM}}_{r,i,j})^2 + \epsilon\right)
\right)
$}
\end{equation}
where $\mathbf{\hat{Y}^{DLRTM}}$ represents the radiative fluxes predicted by DLRTM based on atmospheric state inputs, and $\mathbf{Y^{DLRTM}}$ denotes the ground truth radiative fluxes generated by the RRTMG model. The index $r$ runs over the four radiative flux variables (SWUFLX, SWDFLX, LWUFLX, LWDFLX) across all 13 pressure levels. 
%This loss function ensures that DLRTM accurately reproduces the radiative transfer calculations that would be performed by the more computationally expensive RRTMG model. The constant $\epsilon$ (set to $10^{-6}$) prevents numerical instability when gradients approach zero.
\subsection{FuXi-RTM}
\textbf{Model designs.} FuXi-RTM integrates data-driven weather forecasts with physics-aware constraints by combining two core components: a trainable primary forecasting model (FuXi \cite{chen2023fuxi}) and a pre-trained DLRTM, as illustrated in Fig.\ref{FuXi-RTM_model}. Given sequential atmospheric states $\mathbf{X}_{t-1}$ and $\mathbf{X}_t$, the primary model predicts the next state $\widehat{\mathbf{X}}_{t+1}$, while DLRTM serves as a differentiable physics regularizer that enforces radiative transfer consistency and outputs the coorsponding radiative flux $\widehat{\mathbf{Y}}_{t+1}^\mathbf{DLRTM}$.
%DLRTM is designed to emulate the physics of radiative transfer processes, operating as a surrogate for traditional radiation parameterization schemes. 
During FuXi-RTM's training, DLRTM's parameters remain frozen while it processes outputs from the primary model to generate radiative fluxes. These fluxes are then incorporated into the loss function, creating a physics-guided training signal that enhances forecast accuracy while maintaining computational efficiency.
\begin{figure}
    \centering
    \includegraphics[width=0.5\textwidth]{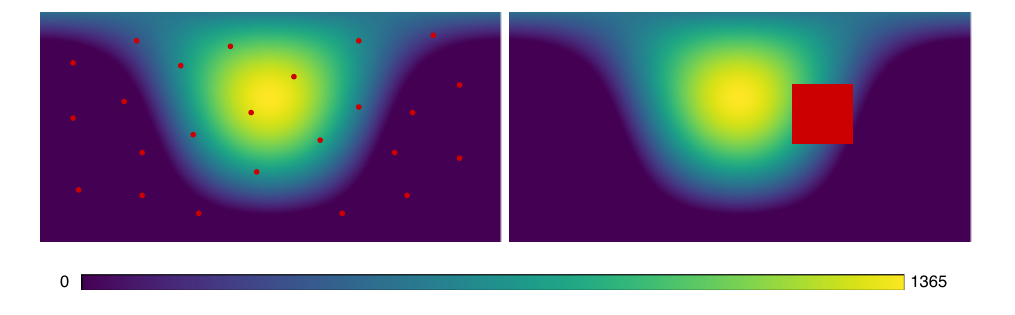}
    \vspace{-20pt}
    \caption{\textbf{Visualized sampling strategy.} Left: Global random sampling. Right: The proposed SRC sampling. The dark regions (0 values) indicating areas without direct sunlight, which are excluded as potential SRC center points. Red points or box represents the sampling locations.}
    % The white area represents the sunlit region, and the red box indicate the random sampled position during one iteration.
    \label{img:1}
\end{figure}
Our implementation focuses specifically on surface-level shortwave upward and downward fluxes (SWUFLX and SWDFLX), as these have shown the greatest influence on forecast improvement. During training, we propose a novel sunlit region-centered (SRC) sampling strategy for loss calculation, which dynamically selects points from a $250\times250$ grid centered on randomly chosen sunlit locations, as illustrated in Fig.\ref{img:1}. This spatial sampling strategy captures meaningful radiative interactions in areas where solar radiation is most impactful.\\%%%%
\textbf{Loss designs.} The training process for FuXi-RTM balances direct forecasting accuracy with physical consistency through a composite loss function. To minimize discrepancies between model outputs and ground truth, we employ a latitude-weighted Charbonnier L1 loss~\cite{Charbonnier1994}:
\begin{equation}
\resizebox{0.48\textwidth}{!}{$
\textrm{L}_{forecast}=
\frac{1}{\textrm{C}\times{\textrm{H}}\times{\textrm{W}}}
\displaystyle\sum_{c=1}^\textrm{C}
\displaystyle\sum_{i=1}^\textrm{H}
\displaystyle\sum_{j=1}^\textrm{W}
\alpha_i
(\sqrt{(\mathbf{\widehat{X}}_{c,i,j}-\mathbf{X}_{c,i,j})^2 + \epsilon^2})
\label{loss_1}
$}
\end{equation}
where $\hat{X}$ and $X$ represent FuXi forecast values and the ground truth respectively. The term $\alpha_i = H \times \frac{\cos\Phi_i}{\sum_{i=1}^{H}\cos\Phi_i}$ is a latitude-specific weighting factor that accounts for the varying grid cell areas at different latitudes, ensuring proper global representation.\\
Additionally, to enforce radiative transfer consistency, we incorporate a physics regularization term:
\begin{equation}
\resizebox{0.46\textwidth}{!}{$
\textrm{L}_{reg}=
\frac{1}{\textrm{R'}\times{\textrm{H'}}\times{\textrm{W'}}}
\displaystyle\sum_{r=1}^\textrm{R'}
\displaystyle\sum_{i=1}^\textrm{H'}
\displaystyle\sum_{j=1}^\textrm{W'}
\alpha_i
(\lambda \sqrt{(\mathbf{\widehat{Y}^{DLRTM}}_{r,i,j}-\mathbf{Y}^{\mathbf{DLRTM}}_{r,i,j})^2 + \epsilon^2})
\label{loss_2}
$}
\end{equation}
where $\textrm{Y}^{\textrm{DLRTM}}$ and $\hat{\textrm{Y}}^{\textrm{DLRTM}}$ denote radiative fluxes generated using ERA5 and FuXi forecasts respectively. 
Here, $H'$ and $W'$ represent the dimensions of the 250×250 grid sampled from sunlit regions, and $\textrm{R'}$ represents the surface-level shortwave fluxes. The parameter $\lambda = 10^{-3}$ balances the contribution of radiative physics constraints against direct meteorological variable prediction. \\
The combined loss function $\textrm{L}_{total} = \textrm{L}_{forecast} + \textrm{L}_{reg}$ drives the model to simultaneously optimize for forecast accuracy and physical consistency, resulting in predictions that better respect the underlying physical processes governing atmospheric dynamics.
\section{Experiment}
We evaluate our approach through experiments measuring forecast accuracy and physical consistency. This section presents our experimental configuration, performance comparisons with the baseline model, ablation studies to validate design choices, and analysis of physical conservation properties.
\subsection{Experimental Setup}
\subsubsection{Dataset Configuration}
We conduct experiments using the ERA5, which represents the state-of-the-art in global atmospheric reanalysis. The model training follows a rigorous temporal split: we use 15 years of data (2002-2016) for training, 1 year (2017) for validation, and a comprehensive 5-year period (2018-2022) for testing. For the test period, forecasts are initialized twice daily at 00:00 UTC and 12:00 UTC, generating predictions at 6-hour intervals up to a 10-day horizon. This extensive temporal coverage ensures robust evaluation across diverse atmospheric conditions and seasonal variations.
\subsubsection{Implementation Details}
Since ERA5 does not include radiative fluxes, we first use the the RRTMG to generate two years (2017-2018) of radiative fluxes based on other ERA5 variables. Using this data, we train our DLRTM surrogate model on NVIDIA A100 GPUs for 5 epochs with a batch size of 41,529. The DLRTM employs the PyTorch framework \cite{Paszke2017} with the AdamW \cite{kingma2017adam,loshchilov2017decoupled} optimizer, configured with $\beta_1 = 0.9$, $\beta_2 = 0.999$, and a cosine annealing learning rate schedule \cite{loshchilov2017sgdr} that decays from $10^{-3}$ to $10^{-8}$. The training set consists of one year of 6-hourly data from 2017, while the validation set includes one year of 6-hourly data from 2018. \\
For the main FuXi-RTM model, we train on a cluster of 4 NVIDIA H100 GPUs for 60,000 iterations with a batch size of 1 per GPU, requiring approximately 81 hours to complete. We use the AdamW optimizer with $\beta_1=0.9$, $\beta_2=0.95$, an initial learning rate of $2.5\times10^{-4}$, and a weight decay coefficient of 0.1.
\begin{figure*}[ht]
\centering
\captionsetup{skip=3pt}  % 局部调整当前图的 caption 间距
\includegraphics[width=1.0\textwidth]{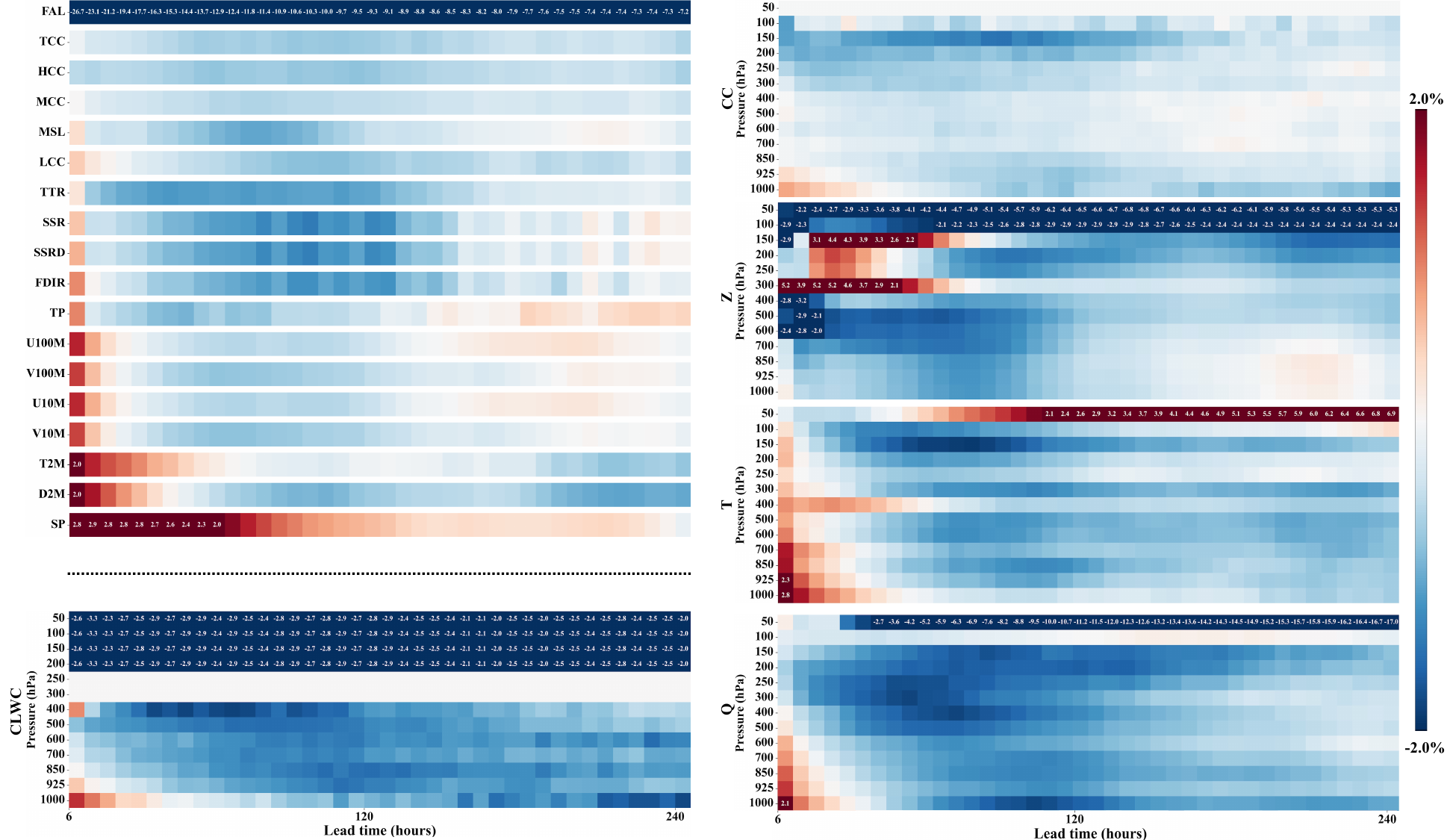}
% \vspace{-70pt}
\caption{\textbf{Scorecard of nRMSE differences in globally-averaged, latitude-weighted RMSE between FuXi-base and FuXi-RTM}. Each subplot corresponds to one of the FuXi direct output variable: 18 surface variables and 5 upper-air variables. For upper-air variables, the rows of each heatmap represent 13 pressure levels. The columns correspond to 40 forecast lead times at 6-hour intervals, spanning from 6 hours to 10 days. The color of each cell indicates the nRMSE differences, with \textbf{blue} denoting negative values (FuXi-RTM outperforms FuXi-base) and red indicating positive values (FuXi-base outperforms FuXi-RTM). The nRMSE difference ranges between -2 and 2, with numeric values overlaid on cells that fall outside this range.}.
\label{scorecard_all}
\end{figure*}
\begin{figure}[ht] %h
\centering
\captionsetup{skip=3pt}  % 局部调整当前图的 caption 间距
\includegraphics[width=0.50\textwidth]{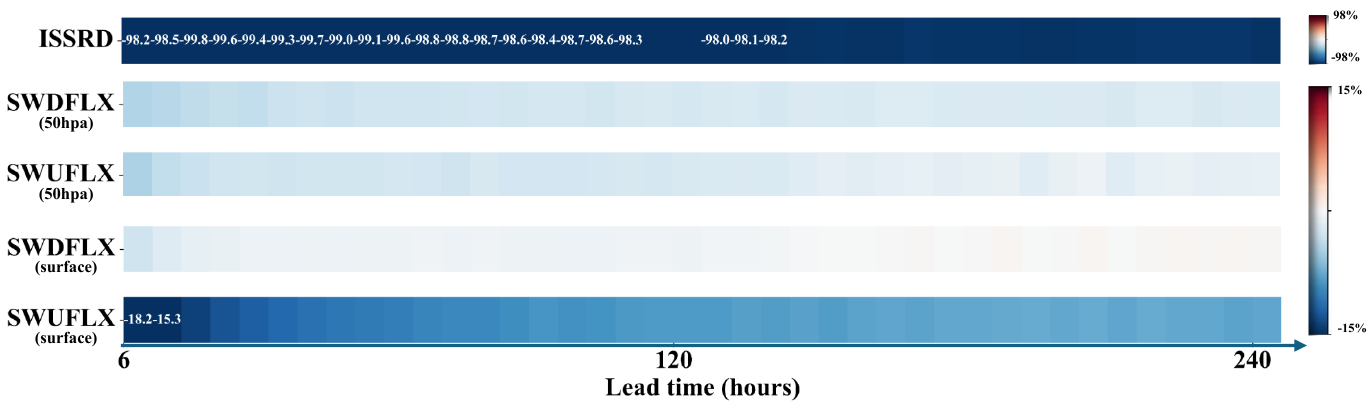}
\caption{\textbf{Scorecard of nRMSE differences in globally-averaged, latitude-weighted RTM RMSE between FuXi-base and FuXi-RTM}. Each subplot corresponds to one of the DLRTM output variable: ISSRD, SWDFLX and SWUFLX at the surface level and 50 hPa. \textbf{Blue} denoting negative values (FuXi-RTM outperforms FuXi-base).}.
\label{scorecard_dlrtm}
\end{figure}
\subsubsection{Metrics}
Unlike conventional video generation tasks that emphasize diversity in predicted futures, weather forecasting has a unique ground truth for each future state, making deterministic evaluation possible and necessary. Following standard practices in operational weather forecast evaluation\cite{chen2023fuxi, ECMWF2021}, we use latitude-weighted root mean square error (RMSE) and normalized RMSE (nRMSE=$\frac{\text{RMSE}-\text{RMSE}_{baseline}}{\text{RMSE}_{baseline}}\times100\%$) as our primary evaluation metrics to account for the varying grid cell sizes across different latitudes.\\
%To better distinguish forecast performance between models with minor differences, the normalized $\textrm{RMSE}$ difference between model A and baseline model B is calculated as \((\textrm{RMSE}_A-\textrm{RMSE}_B)/\textrm{RMSE}_B\times100\%\).A negative RMSE difference indicate that model A outperforms model B. In this study, the baseline model is FuXi-base.
Additionally, to evaluate the physical consistency of radiation predictions, we introduce instantaneous surface net solar radiation downwards (ISSRD) as a specific metric. ISSRD represents the effective solar energy received at Earth's surface and is calculated as:
\begin{equation}
\resizebox{0.4\textwidth}{!}{$
\textrm{ISSRD}_{i,j}=
\frac{\textrm{SWDFLX}_{i,j}^{surface}-\textrm{SWUFLX}_{i,j}^{surface}}{1-\textrm{FAL}_{i,j}}
\label{issrd}
$}
\end{equation}
where $\textrm{SWDFLX}_{i,j}^{surface}$ and $\textrm{SWUFLX}_{i,j}^{surface}$ represent downward and upward shortwave radiative fluxes at surface level, respectively. ISSRD is crucial for weather forecasting, renewable energy planning, and precision agriculture, directly impacting temperature patterns and cloud formation.\\
\subsubsection{Comparative Framework}
We compare our proposed FuXi-RTM against FuXi-base, a pure data-driven weather forecasting model (FuXi) without physical constraints. Additionally, we systematically evaluate FuXi-RTM through controlled ablation studies. We create five model variants, each modifying exactly one aspect of the FuXi-RTM configuration while keeping all other components fixed. These ablations investigate two key design dimensions: (1) \textbf{sampling strategy} - comparing globally random sampling (FuXi-RTM-Random) versus our SRC sampling strategy. (2) \textbf{radiative flux optimization targets} - evaluating four alternatives to surface-level SW fluxes: both SW and LW fluxes across all pressure levels (FuXi-RTM-13level), SW-only fluxes across all levels (FuXi-RTM-13levelSW), net surface SW radiation (FuXi-RTM-GSW), and instantaneous surface net solar radiation (FuXi-RTM-ISSRD).
\begin{figure*}[ht]
\centering
\captionsetup{skip=3pt}  % 局部调整当前图的 caption 间距
\includegraphics[width=1.0\textwidth]{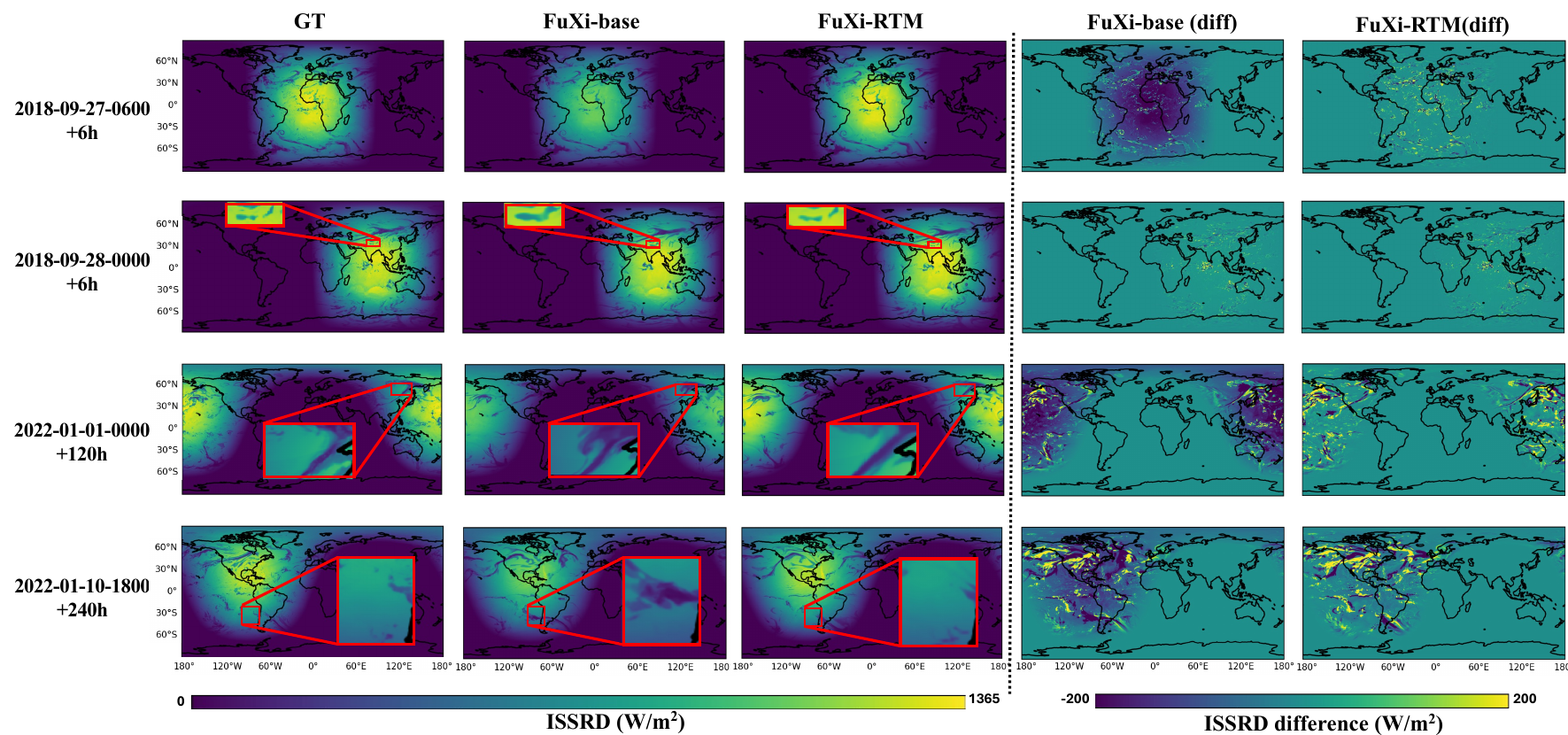}
% \vspace{-70pt}
\caption{\textbf{Snapshot examples of ISSRD.} From left to right: GT (ground truth), FuXi-base (model predictions), FuXi-RTM (model predictions), FuXi-base (diff) (difference between FuXi-base and GT), and FuXi-RTM (diff) (difference between FuXi-RTM and GT). The forecasts are initialized at four different times: 06 UTC on September 27, 2018; 00 UTC on September 28, 2018; 00 UTC on January 1, 2022; and 18 UTC on January 10, 2022. The corresponding forecast horizons are 6, 6, 120, and 240 hours, respectively.}
\label{ISSRD_snapshot}
\end{figure*}
\subsection{Main Results}
Fig.\ref{scorecard_all} presents the normalized differences in globally-averaged, latitude-weighted RMSE between FuXi-base and FuXi-RTM across all 3320 variables and lead time combinations.
Results demonstrate that the hybrid physics-guided architecture, which enforces radiation constraints, significantly improves forecast accuracy.
Specifically, FuXi-RTM outperforms FuXi-base in 88.51\% of all 3320 combinations.
For variables highly related with radiation, such as CC and Q, the percentage of FuXi-RTM's superior performance increases to 95.38\% and 93.46\%, respectively.
Notably, for CLWC, improvements exceed 2\% in nRMSE differences at pressure levels above 200 hPa.
Given the critical role of clouds and radiation in regulating Earth's energy balance, these enhancements underscore FuXi-RTM's potential to advance predictive capabilities in weather and climate forecasting \cite{Liou1986}.
%XH，未来可以对比更多的一些ECMWF HRES和FuXi-RTM的对比结果，包括云量、反射率等。
\begin{table*}[ht] % bt! 顶格
\centering
\small
\caption{\textbf{Ablation studies on different settings.} Performance of FuXi-base, FuXi-RTM and its variants trained under different configurations for critical and classical variables, such as 50-hPa specific humidity (Q50), 500-hPa specific cloud liquid water content (CLWC500). The best and second-best performing variants are highlighted in \textbf{bold} and \underline{underlined}, respectively. Units for each variable are provided in the last row.}
\label{rmse_setup}
\begin{tabular}{lccccccccccccccc}
\hline
\textbf{Model} & \textbf{Q50} & \textbf{Q500} & \textbf{CLWC500} & \textbf{CC150} & \textbf{TCC} & \textbf{TTR} & \textbf{TP} & \textbf{FAL} \\
\hline
FuXi-RTM-Random & 0.1857 & 0.7459 & 0.0226 & 0.1725 & 0.3193 & 159.2536 & 2.4327 & 0.02515 \\
FuXi-RTM-13level & 0.1697 & \underline{0.7409} & 0.0226 & 0.1732 & 0.3184 & 159.2348 & 2.4418 & 0.02313 \\
FuXi-RTM-13levelSW & 0.1762 & 0.7409 & \underline{0.0224} & \underline{0.1714} & \underline{0.3183} & \underline{158.4191} & \underline{2.4200} & 0.02335 \\
FuXi-RTM-GSW & \underline{0.1668} & 0.7567 & 0.0226 & 0.1735 & 0.3207 & 160.9495 & 2.4564 & \textbf{0.02263} \\
FuXi-RTM-ISSRD & 0.1741 & 0.7441 & 0.0226 & 0.1719 & 0.3191 & 159.1379 & 2.4299 & 0.02379 \\
\textbf{FuXi-RTM} & \textbf{0.1546} & \textbf{0.7388} & \textbf{0.0223} & \textbf{0.1705} & \textbf{0.3179} & \textbf{158.1290} & \textbf{2.4092} & \underline{0.02300} \\
FuXi-base & 0.1735 & 0.7453 & 0.0226 & 0.1720 & 0.3196 & 159.1416 & 2.4127 & 0.02553 \\
\hline
Unit & 10\textsuperscript{-3}g/kg & g/kg & 10\textsuperscript{-2}g/kg & 10\textsuperscript{-2}(0 - 1) & (0 - 1) & J/m$^2$ & m & (0 - 1) \\
\hline
\end{tabular}
% }
\label{tab:rmse_comparison}
\end{table*}
However, challenges in training FuXi with DLRTM lead to initial underperformance for certain variables.
For instance, FuXi-RTM trails FuXi-base in predicting Q at 1000 hPa for up to 1.25 days before surpassing it.
The most significant improvements are observed for FAL, with nRMSE differences exceeding 7\% throughout the 10-day forecasts.
This is likely due to FAL's strong coupling with surface SW fluxes, which FuXi-RTM explicitly supervises.
Accurate modeling of FAL is crucial, as it governs the partitioning of Earth’s energy between absorption and reflection \cite{Stephens2015}, directly impacting weather dynamics and long-term climate trends \cite{Dickinson1983,Alex2004}.\\
%XH，值得看一下FuXi-base是不是比ECMWF HRES的反照率预报差，而FuXi-RTM效果更好。我们可以从一些文献中找出，反照率的误差会导致多少预报误差，尤其是长期预报，来凸显他的作用。
\textbf{Radiation Prediction Evaluation.} Fig.\ref{scorecard_dlrtm} illustrates the nRMSE differences for DLRTM output variables, including ISSRD as well as SWDFLX and SWUFLX at the surface level and TOA (50 hPa in this study).
%FuXi-base occasionally produces unphysical outputs, such as ISSRD values below 0 or exceeding 1360 W/$m^2$ (the solar constant, which represents the theoretical maximum ISSRD value for an atmosphere-free Earth with the sun directly overhead).
%Figure \ref{scorecard_dlrtm} further compares the normalized differences in the occurrences of unphysical ISSRD values between FuXi-base and FuXi-RTM across all lead times. 
Our evaluation reveals that FuXi-RTM achieves modest improvements in predicting SWDFLX and SWUFLX at 50 hPa despite only explicitly constraining surface-level shortwave radiation. Furthermore, FuXi-RTM demonstrates substantially smaller RMSE for ISSRD predictions with improvements approaching 100\%, while the performance gains for SWDFLX and SWUFLX remain comparatively modest. Analysis of the ISSRD computation formula \ref{issrd} reveals that this significant improvement is primarily attributable to the higher accuracy of the Fraction of FAL term, which serves as the denominator in the calculation, thereby amplifying the relative enhancement in prediction quality.
%XH, 在这里通过分析SWDFLX和SWUFLX在50hPa的表现，来讨论T为啥在高空变差了。

To gain deeper insights into ISSRD performance, Fig.\ref{ISSRD_snapshot} compares global spatial distribution of ground truth (GT) with the predicted ISSRD from FuXi-base and FuXi-RTM for four randomly selected forecast initialization dates and lead time.
The difference maps between predictions and GT, shown in the right two columns, clearly reveal that FuXi-RTM achieves smaller errors compared to FuXi-base. This improvement is particularly evident in sunlit regions, where FuXi-base exhibits widespread negative ISSRD biases. These spatial patterns align with the nRMSE differences quantified in our earlier analyses, further validating the enhanced performance of physics-constrained approach.\\
\textbf{DLRTM Evaluation.} Beyond accuracy metrics (see supplementary material), we evaluate computational efficiency. DLRTM achieves orders of magnitude speedup over the traditional RRTMG model through parallel batch processing, reducing computation time from 22 minutes (8 CPUs) to approximately 3 seconds (1 H100 GPU) for global ($721\times1440$) grid points.
\subsection{Ablation Study}
\label{comparison_setup}
Tab.\ref{rmse_setup} compares the RMSE of FuXi-base, FuXi-RTM, and its variants trained under different configurations for key meteorological variables.
FuXi-RTM consistently outperforms both its variants and FuXi-base across most variables. %, with the exception of FAL
Notably, all FuXi-RTM variants demonstrate superior overall performance compared to FuXi-base.\\
%Experiments demonstrated that SRC sampling and optimizing surface-level SW fluxes yielded the highest forecast accuracy.
%Models employing our SRC sampling significantly outperformed random global sampling, demonstrating that spatially coherent gradient computation enhances feature learning by preserving local contextual relationships while reducing interference from irrelevant background signals.\\
Our ablation studies reveal several key insights. Models employing our sunlit region-centered sampling significantly outperformed random global sampling (FuXi-RTM-Random), demonstrating that spatially coherent gradient computation enhances feature learning by preserving local contextual relationships while reducing interference from irrelevant background signals. For variants optimizing all pressure levels (FuXi-RTM-13level and FuXi-RTM-13levelSW), we observed diminished performance likely due to information redundancy in the radiative transfer process—surface radiation constraints inherently capture vertical atmospheric interactions through backpropagation, making explicit multi-level supervision unnecessary and potentially counterproductive. The FuXi-RTM-GSW and FuXi-RTM-ISSRD variants, which optimize derived radiation metrics rather than direct fluxes, underperformed compared to FuXi-RTM. This suggests that constraining fundamental radiative components (surface SW fluxes) provides the model with more comprehensive physical information than derived quantities that inherently contain less information about the underlying radiative processes.
%XH，反射率与短波辐射可能具有较强的相关性，但是这个相关性我们怎么表达需要考虑清楚。
\begin{figure}
\centering
\captionsetup{skip=3pt}  % 局部调整当前图的 caption 间距
\includegraphics[width=0.49\textwidth]{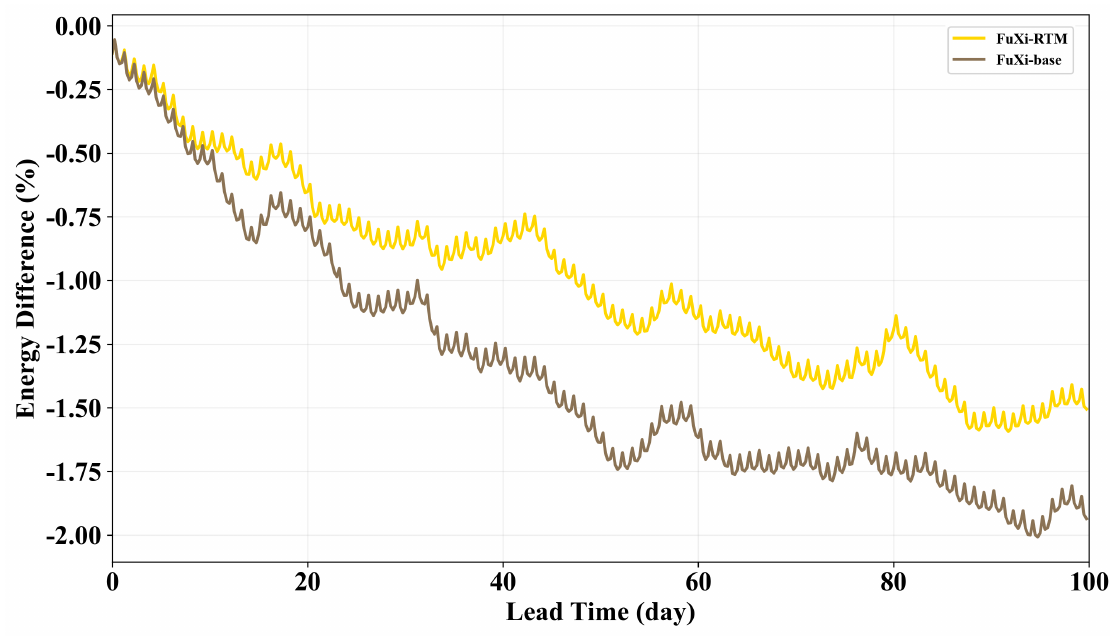}
% \vspace{-70pt}
\caption{\textbf{Verification of global total atmospheric energy conservation.} Normalized differences in global total atmospheric energy of FuXi-RTM (yellow) and FuXi-base (brown) forecasts relative to ERA5-derived reference values. Values represent averages over forecasts initialized at 00 UTC on five random dates in 2018. The horizontal axis represents the forecast duration, and the vertical axis represents the percentage of energy loss. The lower the value on the vertical axis, the more severe the energy loss.}
\label{plot_conservation_energy}
\end{figure}
\subsection{Global total atmospheric energy conservation}
\label{conservation_energy}
To further validate the physical consistency of our approach, we examine the conservation of global total atmospheric energy (detailed formulation in supplementary material). As shown in Fig.\ref{plot_conservation_energy}, FuXi-RTM demonstrates superior energy conservation compared to FuXi-base, with the difference becoming particularly pronounced beyond 10-day forecasts. These enhanced conservation properties directly result from our radiative constraints, confirming FuXi-RTM's improved physical fidelity, which becomes increasingly important for long-term predictions.
\section{Conclusion} 
FuXi-RTM is a hybrid physics-guided deep learning model that innovatively enforces radiation constraints by employing a deep learning surrogate for radiation parameterization.
To the best of our knowledge, FuXi-RTM is the first deep learning-based weather forecasting model to explicitly model physical processes, as opposed to integrating a dynamical core with a NN serving as a holistic parameterization.
Overall, FuXi-RTM demonstrates that incorporating radiation constraints not only enhances forecasts for radiation and clouds but also improves predictions for conventional meteorological variables.
%, such as geopotential and humidity.
Several potential avenues exist for further improving FuXi-RTM.
First, the model currently excludes the u and v wind components at all pressure levels. Scaling up to include the uv wind and additional components could further improve the model's performance, though this would necessitate balancing the trade-off with increased computational resource requirements.
Second, the hybrid physics-guided architecture of FuXi-RTM is adaptable to the incorporation of additional physical processes, such as convection, PBL, and cloud microphysics, into deep learning-based weather forecasting framework.
By addressing the limitations, this hybrid architecture paves the way for next-generation weather forecasting systems that are accurate, efficient, and trustworthy.
%XH, 这个点可能需要一些实验验证，我们可以看看辐射的极端值的预报，加一些极端的指标

% Many applications require accurate forecasts of surface radiation on weather timescales, for example solar energy and UV radiation forecasts.
{
    \small
    \bibliographystyle{ieeenat_fullname}
    \bibliography{ref}
}

% https://github.com/climlab/climlab-rrtmg
% 补充 大气守恒计算公式（附录）、创新点突出简单、高效、无需额外训练直接预测辐射的角色定位、确认物理约束自然视频生成的rw。
% 写作技巧总结：
% method不要拖太后让审稿人失去耐心
% 能公式就不要文字
% 如果有类似工作，先一气呵成写完，再说Note that there is a rencent/concurrent work。。。或者就是总结区别陈述
% 有事实根据的论述要引文，没有事实根据的可以用it is plausible that或者we guess
% 题目不要太长也不要太短
% 陈述效果的时候要客观，不然会被抨击
% 在reference缩短篇幅的时候，可以缩写ICLR
\end{document}